\documentclass[aps,prl,twocolumn,groupedaddress]{revtex4}
\usepackage{amsmath}
\usepackage{graphicx}
\usepackage{subfigure}
\usepackage{epstopdf}

\newcommand{\beq}{\begin{equation}}
\newcommand{\eeq}{\end{equation}}

\usepackage{color}

\usepackage{bm}
\usepackage{bbm}
\usepackage{epsfig}

\begin{document}

%\preprint{}

\title{Dijet resonance from leptophobic ${\bm Z^{'}}$ and \\
light baryonic cold dark matter }
%\\ in (non)supersymmetric $U(1)_B \times U(1)_L$ model}


\author{P. Ko}
\email[]{pko@kias.re.kr}
\affiliation{School of Physics, KIAS, Seoul 130-722, Korea}

\author{Yuji Omura}
\email[]{omura@kias.re.kr}
\affiliation{School of Physics, KIAS, Seoul 130-722, Korea}

\author{Chaehyun Yu}
\email[]{chyu@kias.re.kr}
\affiliation{School of Physics, KIAS, Seoul 130-722, Korea}


%\date{\today}

\begin{abstract}
In light of the recent CDF report on the excess in the $Wjj$ channel, 
we analyze (non)supersymmetric $U(1)_B \times U(1)_L$ model,
interpreting the dijet peak as a leptophobic $U(1)_B$ gauge boson. 
If this excess is confirmed, it has an interesting implication for
the baryonic cold dark matter (CDM) in the model
: there should be light 
CDM with a few GeV mass, and direct detection cross section at the level
of a few $\times 10^{-2}$ pb. 

%Also $Z_B Z_B$ channel can contribute to the ultra massive jets in the 
%dijet mass range $130$ GeV $< m_{jj} < 210$ GeV.
\end{abstract}

%\keywords{}

\maketitle

\section{Introduction}
%\subsection{}
%\subsubsection{}

Recently the CDF Collaboration reported an excess in the $Wjj$ channel, 
with a broad peak in the dijet with mass around 120 -- 160 GeV 
\cite{Aaltonen:2011mk}.  
There is no evidence for enhancement in the $l \nu j j $ invariant mass 
spectrum, so that the excess is less likely to be from a single 
$s$-channel resonance in $q\bar{q}$ annihilation. Also the dijets 
in the final state are not dominantly $b$-flavored.  
It would be amusing to speculate what would be the underlying 
physics for this excess.  
A simple interpretation of this excess would be to assume a new 
spin-1 particle with mass around 140 GeV. In order to avoid the strong
constraints from Drell-Yan production, this new spin-1 object better 
be leptophobic, or its leptonic branching ratio should be very small.
There appeared a number of papers which discuss this excess in various 
contexts: \cite{Buckley:2011vc,Yu:2011cw,Eichten:2011sh,Kilic:2011sr,%
Wang:2011uq,Cheung:2011zt,AguilarSaavedra:2011zy,He:2011ss,Wang:2011taa,%
Sato:2011ui,Nelson:2011us,Dobrescu:2011px,Sullivan:2011hu}.

Very recently, the D0 Collaboration also reported their analysis
on the $W+jj$ production with similar experimental cuts 
to the CDF's ones~\cite{Abazov:2011af}. Unlike the CDF results,
the D0 Collaboration did not observe any excess on the dijet.
However the previous results by the CDF Collaboration are consistent 
with  the analysis with larger data sample of an integrated 
luminosity of $7.3$ fb$^{-1}$ at the CDF~\cite{CDFwjj2}.
Up to now two analyses are in conflict with each other and 
we could not exclude the possibility that both results are statistical
fluctuation. Eventually this issue should be settled down by more data
analysis at the Tevatron and the LHC.

In this Letter, we consider leptophobic $Z^{'} \equiv Z_B$, 
associated with gauged $U(1)_B$, based on our recent model 
\cite{Ko:2010at}.
A nonsupersymmetric anomaly-free $U(1)_B \times U(1)_L$ model 
was constructed in Ref.~\cite{FileviezPerez:2010gw}, and      
the model was extended  to supersymmetric one 
by two of the present authors  \cite{Ko:2010at}. 
(For earlier studies on gauged $U(1)_B$ model, we refer to  
Refs.~\cite{Mangano:1996ab,Drees:1998rz} and references therein.)
The supersymmetric (SUSY) version \cite{Ko:2010at} has both 
baryonic and leptonic cold dark matter (CDM), 
in addition to the lightest neutralino CDM,
thereby the dark matter sector having very rich structure.  
In these models, the baryonic gauge boson $Z_B$ has a universal 
coupling to the SM quarks, and three times larger to the new mirror 
quarks which are introduced to cancel anomalies. 
This model has a natural color-singlet baryonic CDM with $U(1)_B$ 
charge twice larger than the SM quarks.  Therefore $Z_B$ can decay 
into a pair of baryonic CDM's, if the CDM is lighter than half the 
$Z_B$ mass $M_{Z_B}/2$. 
The new mirror quarks could have constraints from search for 
the 4th generation fermions. 
The masses of exotic quarks should be more than $\sim$ 300 GeV
~\cite{Alwall:2010jc}, which requires very large Yukawa couplings
leading to Landau poles at a low scale.  

We interpret the excess reported by the CDF Collaboration in the $l\nu jj$ 
channel as $p \bar{p} \rightarrow W Z_B \rightarrow (l \nu) ( j j )$. 
Then the CDF data provide informations on $M_{Z_B}$ and the $U(1)_B$ gauge 
coupling $g_B (\equiv \sqrt{4 \pi \alpha_B})$. 
These informations can be used to study the thermal relic density 
and the direct detection cross section of baryonic CDM 
in gauged $U(1)_B \times U(1)_L$ model, as well as other collider 
signatures such as $\gamma Z_B, Z Z_B , Z_B Z_B$. 
We find that the fermionic CDM in supersymmetric $U(1)_B \times U(1)_L$ 
model  can be as light as $\sim 5$ GeV, 
with $\sigma_{SI} \sim$ (a few) $\times 10^{-2}$ pb, 
which is somewhat larger than the CoGeNT~\cite{Aalseth:2010vx} and 
DAMA~\cite{Bernabei:2010mq} signal region.
%{\color{blue} 
%but has not been explored by any direct detection experiments 
%of light cold dark matter. 
%}

\section{Gauged $\bm{U(1)_B \times U(1)_L}$ model}

It is well known that $U(1)_B$ is anomalous within the standard 
model (SM), and one has to 
introduce new matter fields in order to %gauge $U(1)_B$ symmetry and 
cancel all the gauge anomalies when 
we introduce $U(1)_B$ gauge boson which is leptophobic.  
Recently a simple model was proposed where one family of mirror fermions 
with baryon number $Q_B = 1$ were introduced for this purpose. 
Then another new complex scalar $X_B$ with $Q_B = 2/3$ 
was introduced in order to make the heavy mirror fermions decay
through the following Yukawa interactions,
\begin{equation}
{\cal L}_Y=-\lambda_{Qi} X_B \overline{Q'}Q_i 
-\lambda_{Di} X^{\dagger}_B \overline{D'}D_i
-\lambda_{Ui} X^{\dagger}_B \overline{U'}U_i +h.c., 
\end{equation}
where $Q'$, $D'$, and $U'$ are the extra mirror quarks required 
for the anomaly-free conditions and $\lambda_i$'s are the corresponding
Yukawa couplings.
This new scalar $X_B$ carrying baryon charge becomes stable due to 
accidental symmetry, and becomes a good candidate for baryonic CDM 
of the universe \cite{FileviezPerez:2010gw}. 
In the supersymmetric $U(1)_B \times U(1)_L$ model,
new chiral superfields $X_L$ and $\overline{X_L}$ were introduced, 
lighter of which (either bosonic or fermionic) can make leptonic
CDM \cite{Ko:2010at}.  
Also the superpartner of $X_B$,  Dirac fermion $\widetilde{X_B}$, 
can be another candidate for baryonic CDM. 
In addition, SUSY $U(1)_B \times U(1)_L$ model
has ordinary lightest neutralino as a possible candidate for CDM. 
Therefore SUSY $U(1)_B \times U(1)_L$ model has a rich structure in dark 
matter sector.  In this Letter, we concentrate on $U(1)_B$ part only, so
we will drop $U(1)_L$ model from now on. 

If we consider the broad peak in dijet invariant mass reported by the 
CDF Collaboration as a leptophobic $Z_B$ decaying to $q\bar{q}$
and the bound on $g_B$ from the $pp\to jj$ process 
in the UA2 experiments~\cite{ua2,ua2-2},
we have important piece of informations on our model: namely 
$g_B \sim 0.8$ and $M_{Z_B} \sim 140$ GeV \cite{Yu:2011cw}. 
Then we can study more phenomenology of gauged $U(1)_B$ model, 
both supersymmetric and nonsupersymmetric ones. 
In particular, the cold dark matter sector can be  constrained from the 
informations on $g_B$ and $M_{Z_B}$ from the CDF data, 
thermal relic density, and the upper bounds on the direct detection rates.

The complete $U(1)_B \times U(1)_L$ model has the mirror fermions 
and their superpartners, and they can also affect the dark matter physics 
through Yukawa couplings.  
In this Letter, we assume the Yukawa couplings involving mirror particles 
are very small in order 
to reduce the number of unknown parameters and simplify the analysis. 
%In this limit, the $Z_B$ gauge boson is the most important one from 
%the $U(1)_B$ model, which is less dependent on the details of the model. 
Then $U(1)_B$ gauge interaction is the only new relevant one, and 
the mirror fermions do not affect significantly the CDM physics 
we describe here.  Including Yukawa couplings to the mirror particles 
will be another important 
subject for further study. %, beyond the scope of this letter.
%(See, for example, Ref.~\cite{}.)

\section{CDF data on $\bm{W+jj}$}

We assume that the CDF data on $W+jj$ are due to the $W Z_B$ boson 
production with $M_{Z_B} \sim 140$ GeV and $g_B \sim 0.8$.
Then, the $Z_B$ could be identified in other diboson channels
like the $Z Z_B$, $\gamma Z_B$ and $Z_B Z_B$ production processes
if the SM backgrounds can be controlled  \cite{Cheung:2011zt}.
Up to the now, there is no significant excess in the $Z+ jj$ events so far
\cite{Aaltonen:2011mk}, and it remains to be seen what happens 
in this channel in the forthcoming analysis from the Tevatron and the LHC.
%updated CDF analysis as well as in .

\begin{figure}
\begin{center}
\begin{tabular}{c}
\epsfig{file=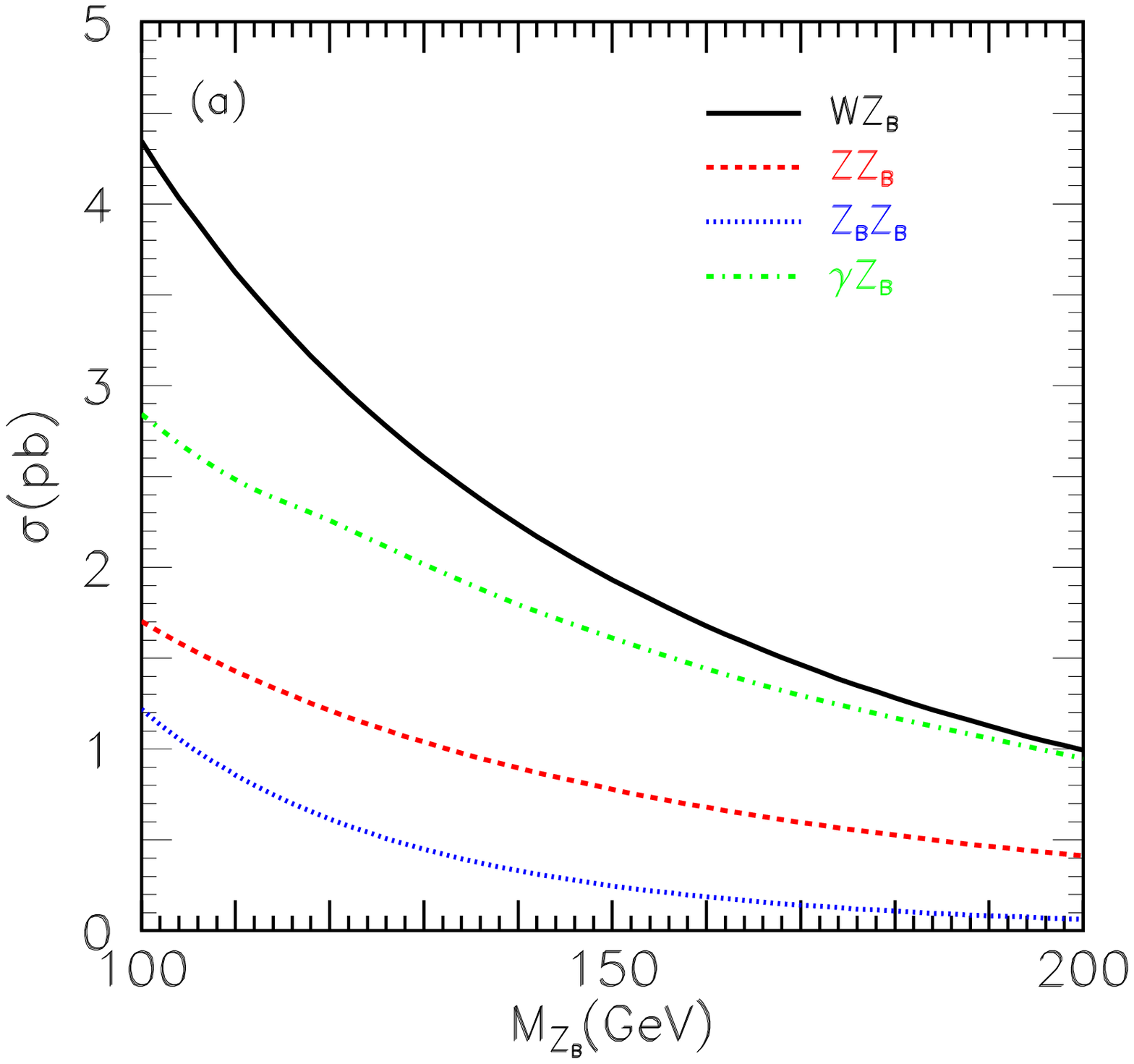,width=0.45\textwidth}
\\
\epsfig{file=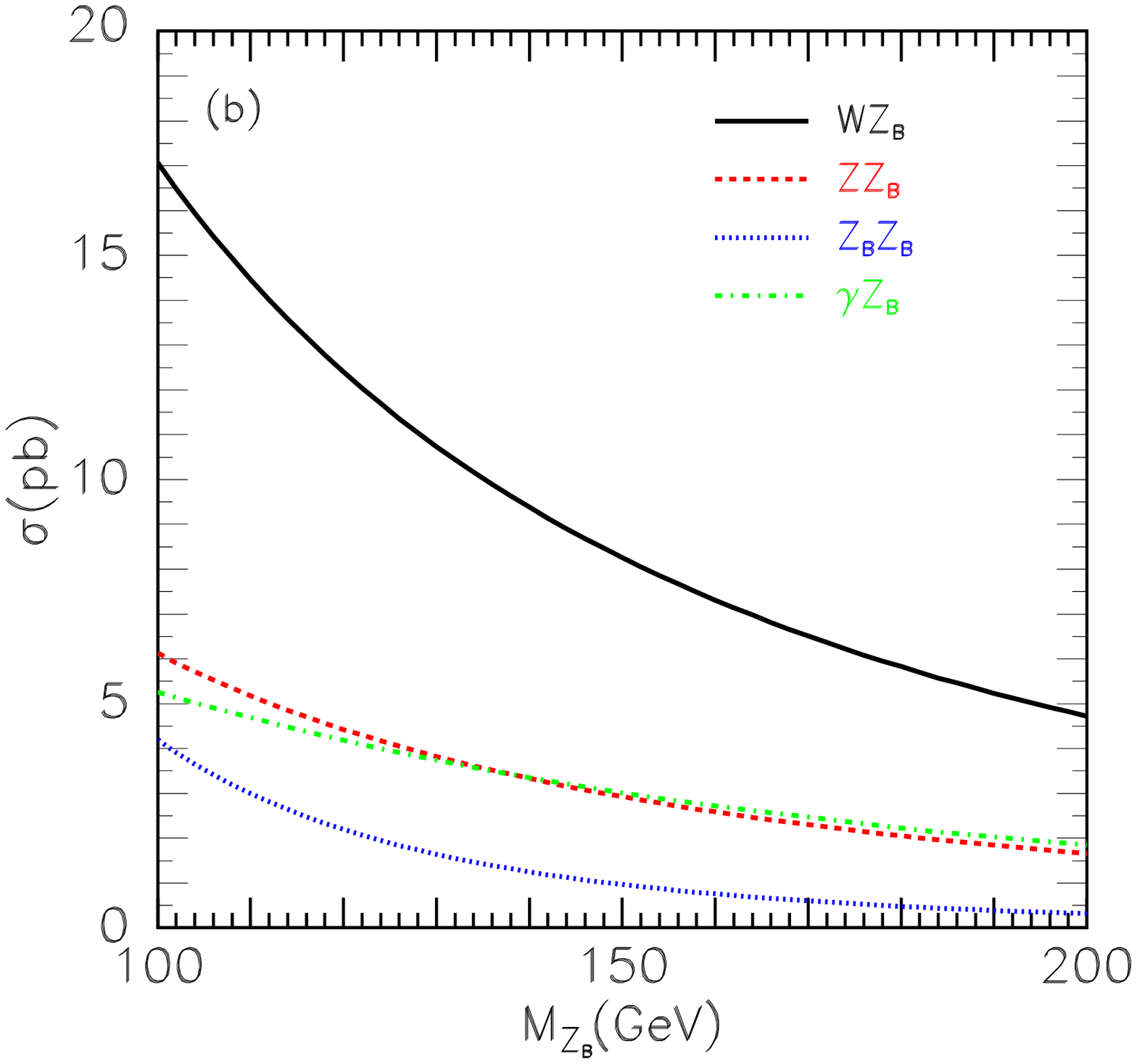,width=0.45\textwidth}
%\\[-8.5ex]
\end{tabular}
\caption{\label{fig:crosssection} 
Production cross sections for $W Z_B$,  $Z Z_B$, $Z_B Z_B$
and $\gamma Z_B$  (a)  at the Tevatron ($\sqrt{s}=1.96$ TeV) and 
(b) at the LHC ($\sqrt{s}=7$ TeV) 
as functions of $Z_B$ mass for $g_B = 0.8$. 
For the $\gamma Z_B$ mode, we apply  the photon transverse-momentum 
and pseudorapidity cuts, $p_T^\gamma > 30$ GeV and $|\eta^\gamma|<1.1$.
}
\end{center}
\end{figure}

In Fig.~\ref{fig:crosssection} (a) and (b), we show the cross sections
for the $W Z_B$, $Z Z_B$, $\gamma Z_B$ and  $Z_B Z_B$ productions  
at the Tevatron with the center-of-momentum energy $\sqrt{s}=1.96$ TeV 
and at the LHC with $\sqrt{s}=7$ TeV, respectively, 
as functions of the $Z_B$ mass $M_{Z_B}$ with $g_B=0.8$ 
imposing the UA2 bound \cite{ua2,ua2-2}.
For the $\gamma Z_B$ production, we impose the photon transverse-momentum 
cut $p_T^\gamma > 30$ GeV and the photon pseudorapidity cut
$|\eta^\gamma|< 1.1$, which are consistent with the experiments
at the Tevatron~\cite{CDFnote}.
The cross sections for other $g_B$ values can be easily scaled by $(g_B/0.8)^2$
for the $W Z_B$, $Z Z_B$ and $\gamma Z_B$ channels
and by $(g_B/0.8)^4$ for the $Z_B Z_B$ channel, respectively.
For $M_{Z_B}=140$ GeV and $g_B=0.8$, we find that $\sigma(W Z_B) = 2.2$ pb
at the Tevatron,
which is about half the cross section for the $W+jj$ excess at CDF
with an assumption on the hypothesized narrow Gaussian contribution.
In order to fit the cross section to the CDF excess, we can require 
a larger coupling with smaller $Z_B$ mass.   Or  the current CDF data
could be an upper fluctuation.  This issue could be resolved in the near future
with more data accumulated and analyzed. 
In the other diboson productions, we find that
$\sigma(Z Z_B)=0.90$ pb, $\sigma(Z_B Z_B)= 0.33$ pb and 
$\sigma(\gamma Z_B)=1.8$ pb at the Tevatron for $M_{Z_B}=140$ GeV
and $g_B=0.8$, respectively. 
At the LHC, we expect that $\sigma(W Z_B)= 9.4$ pb,
$\sigma(Z Z_B)=3.3$ pb, $\sigma(Z_B Z_B)=1.3$ pb and
$\sigma(\gamma Z_B)=3.3$ pb, respectively. 
In order to make definite conclusion about the possibility to find the 
$Z_B$ boson in the diboson channels at the Tevatron or at the LHC, 
we need more thorough study  on the signal-to-background ratio with 
the detector simulation, which is out of scope of this work. 

%A remark is in order about $Z_B Z_B$ production at the Tevatron. 
%It is amusing to speculate that the $Z_B Z_B$ production could contribute 
%to the ultra massive boosted jets observed at CDF in the mass range 
%$130$ GeV $< m_{j} < 210$ GeV at the level of $\sim O(10)$ fb
%(Ref.~\cite{Eshel:2011vs} and references therein)  
%if we use smaller $g_B$, assuming that the $Wjj$ excess reported by CDF 
%is an upward fluctuation. This mode deserves further study \cite{e6}.

%\begin{widetext}
%\end{widetext}

\section{Baryonic cold dark matter}

The CDF data on $W +jj$ events can be accommodated with leptophobic
$Z_B$ gauge boson, if $g_B \sim 0.8$ and $M_{Z_B} \sim 140$  GeV. 
If we take this value in the gauged $U(1)_B \times U(1)_L$ model, 
one can constrain the dark matter sector more or less from the WMAP 
measurement of thermal relic density of CDM and upper bounds from 
direct detection experiments. 

For nonsupersymmetric $U(1)_B$ model, a baryonic complex scalar $X_B$
can make a good CDM candidate. Neglecting its Yukawa couplings to the 
mirror fermions $Q^{'}, u^{'}$ and $d^{'}$, we can calculate thermal relic
density from $X_B \overline{X_B} \rightarrow Z_B \rightarrow$ SM particles.
It turns out that thermal relic density of bosonic $X_B$ is too large, unless
$m_{X_B} \simeq M_{Z_B}/2 \sim 70$ GeV (the $s$-channel resonance 
annihilation into the SM quarks). 
In order to achieve small enough relic density consistent with the WMAP
data without using the $s$-channel resonance annihilation, 
other channels involving mirror fermions and their superpartners 
need to be considered.   %as recently discussed in Ref.~\cite{Buckley:2011vs}.
Also, if the CDF dijet excess becomes less prominent in the future  
and $g_B$ becomes smaller, we have to invoke Yukawa couplings to 
mirror fermions in order to get the correct thermal relic density.

\begin{figure}
\includegraphics[width=7cm]{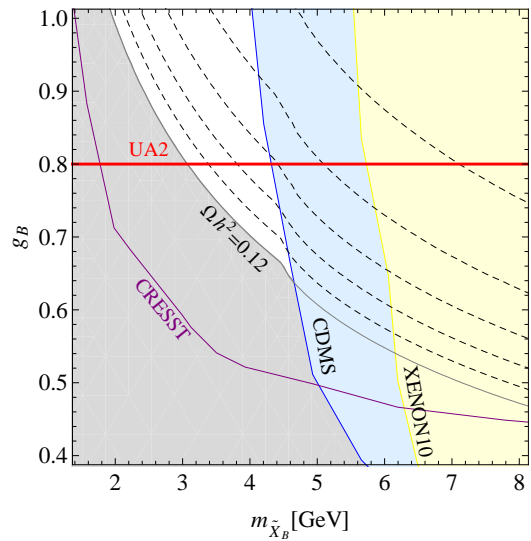}
\caption{\label{fig:XBDM} 
Relic density of  baryonic CDM, $\widetilde{X_B}$.
The gray region corresponds to $\Omega_{\widetilde{X_B}} h^2 \geq 0.12$ 
and each dashed line is for 
$\Omega_{\widetilde{X_B}} h^2 = 0.10,~0.08,~0.06,~0.04,~0.02$ 
from bottom to top.
The yellow region is excluded by XENON10 (90 \% C.L.) and 
the blue is CDMS (90 \% C.L.)~\cite{Gondolo}.
%{\color{blue} with zero 
%${\cal L}_{eff}$ below $3.9$ keVnr. 
%}
The red line is the UA2 bound on $g_B \leq 0.8$. }
\end{figure}

For supersymmetric case, Dirac fermion $\widetilde{X_B}$ and its antiparticle 
carrying $Q_B = \pm 2/3$ can be good CDM candidates \cite{Ko:2010at}, 
because the annihilation cross section has $S$-wave contribution.  
%In this letter, we consider a scenario where a light CDM is the Dirac fermion 
%$\widetilde{X_B}$, which corresponds to the superpartner of $X_B$  and 
%carries baryon number $Q_B = 2/3$ }. 
In Fig.~\ref{fig:XBDM}, we show the contour plots for thermal relic density 
($\Omega_{\widetilde{X_B}} h^2$) in the $(m_{\widetilde{X_B}}, g_B)$ plane. 
%{\color{blue}
%The yellow region is excluded by XENON10 (90 \% C.L.) with zero scintillation 
%efficiency factor ${\cal L}_{eff}$ below $3.9$ keVnr~\cite{Angle:2007uj}. }
%The $U(1)_B$ coupling $g_B$ may be 
%constrained by collider experiments, 
%such as UA2 and Tevatron \cite{Jung:2011ua}.
%In our model, $Z_B$ can annihilate to CDM, so that we need 
%more careful analysis 
%\cite{e6}.  
%In any case, 
There remains a small corner of parameter space 
with $m_{\widetilde{X_B}} \sim 4 - 6$ GeV and 
$g_B < 0.8$ (the red line) which could be safe against the UA2 bound.  
In this region of parameter space, the direct detection cross section is around 
$\sigma_{SI} \sim 0.01-0.05$pb, which is slightly above the CRESST bound, 
$\sigma_{SI} \lesssim O(10^{-3})$pb~\cite{Gondolo}. 
%There seems to be still ambiguity in each experimental results,
%{\color{red} except for the CRESST experiment. 
There are several CDM candidates in our model,
so that $\widetilde{X_B}$ could be subdominant.
If $m_{\widetilde{X_B}}$ is heavier or the Yukawa contribution is large enough  
to reduce the relic density of $\widetilde{X_B}$, the upper bound 
of $\sigma_{SI}$ could be enhanced by the factor 
$ (0.11/ \Omega_{\widetilde{X_B}}h^2)$.
However, we may face the stronger bound from collider experiments 
in the scenario with light CDM, as we discuss in the below.
%{\color{blue} 
%It is highly desirable to improve direct detection in the mass range of 
%a few GeV with $\sigma_{\rm SI} \sim O(10^{-2})$ pb. 
%}
%\section{$X_B$ CDM}

\section{Further collider signatures}

For non-SUSY $U(1)_B$ model, the bosonic baryonic CDM $X_B$ 
has mass close to $M_{Z_B}/2$ if we fix $M_{Z_B} \sim 140$ GeV and 
$g_B \sim 0.8$ in order to explain the CDF W+jj excess. 
For these parameter values, the invisible decay width of $Z_B 
\rightarrow X_B X_B^\dagger$ will be negligible.  
If $g_B$ turns out smaller or if one would like to use other channels 
rather than the $s$-channel annihilation of the dark matter pair 
through $Z_B$, it would be possible to have light bosonic $X_B$ without 
conflict with the direct detection bounds.  In this case, the invisible decay 
$Z_B \rightarrow X_B X_B^\dagger$ could be possible.
This would help to study the diboson productions with at least one $Z_B$.
However, this possibility depends on  parameters other than 
$M_{Z_B}$ and $g_B$, and we do not consider this case further in this Letter.

\begin{figure}
\begin{center}
\epsfig{file=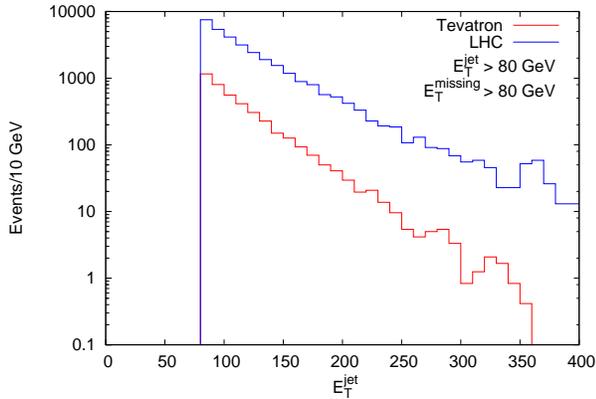,width=0.45\textwidth}
\caption{\label{fig:monojet} 
The distribution of the number of the jet in the monojet production 
at the Tevatron (red line) and LHC (blue line)
as a function of the transverse energy of the jet for $g_B=0.8$ and 
the dark matter mass $m_X=5$ GeV with the integrated luminosity of
1 fb$^{-1}$.
}
\end{center}
\end{figure}

For SUSY version, the fermionic baryonic CDM could be light so that 
the invisible decay mode can have 
$B( Z_B \rightarrow \widetilde{X_B} \overline{\widetilde{X_B}}) 
\approx 21 \%$.  
%(an invisible decay of $Z_B$)  $\approx 20 \%$ branching ratio,
Then high $p_T$ monojet (or single photon) with large missing $E_T$ from 
$q\bar{q} \rightarrow g Z_B~ ({\rm or}~ \gamma Z_B)$ or 
$q (\bar{q}) g \rightarrow q (\bar{q}) Z_B$ followed by 
$Z_B \rightarrow \widetilde{X_B}\overline{\widetilde{X_B}}$ 
would make clean signatures of our model. 
Note that the missing $E_T$ signature from 
$Z_B \rightarrow \widetilde{X_B}\overline{\widetilde{X_B}}$ 
makes a unique feature of gauged $U(1)_B$ model with light  baryonic 
cold dark matter.
The $q\bar{q^\prime}\rightarrow W Z_B \rightarrow 
(l \nu) (\widetilde{X_B} \overline{\widetilde{X_B}})$ 
channel could lead to a single high $p_T$ charged lepton plus missing $E_T$,
which however would suffer severe background at hadron colliders.
In Fig.~\ref{fig:monojet}, we depict the distribution 
of the number of the jet in the monojet production at the Tevatron and LHC
as a function of the transverse energy of the jet for $g_B=0.8$ and 
the dark matter mass $m_X=5$ GeV. The distribution was generated by
using MADGRAPH \cite{madgraph}.
We applied the cuts on the transverse energy
of the jet ($E_T^{\textrm{jet}}>80$ GeV) and the transverse energy 
of the missing momentum ($E_T^{\textrm{missing}} > 80$ GeV), respectively.
The integrated luminosity is assumed to be 1 fb$^{-1}$. The expected event
numbers are about 4000 and 30,000 at the Tevatron and LHC, respectively.
The cross section for the monojet production could be decreased by a factor of
$g_B^2$ as the $g_B$ is decreased.

If the $M_{Z_B}$ mass is around 140 GeV, then the two jets + missing
energy signals through $e^+ e^- \to q \bar{q} \to q\bar{q} + Z_B$ with
the subsequent decay $Z_B \to \widetilde{X_B} \overline{\widetilde{X_B}}$
at LEP II may give useful constraints on $M_{Z_B}$ and $g_B$ \cite{Drees}.
We find that $\sigma ( q\bar{q} Z_B) \times Br( Z_B)_{\rm invisible} \simeq
2 \times 10^{-5}$ pb for $M_{Z_B} =140$ GeV and $g_B = 0.8$, 
which is out of reach at LEP II. It could be studied at future linear colliders.

The mass of baryonic scalar $S_B$, whose nonzero VEV breaks $U(1)_B$ 
spontaneously, can be as large as a few hundred GeV 
in non-SUSY case, and it will mix with the SM Higgs boson $h_{\rm SM}$. 
$S_B \rightarrow Z_B Z_B \rightarrow 4 q's$, or 
$S_B \rightarrow h_{\rm SM} \rightarrow b\bar{b}, t\bar{t}, WW, ZZ$
depending on $S_B$ mass.  Production of $S_B$ is by $S_B-$strahlung (similarly 
to the Higgs-strahlung), and the production rate will be smaller than the SM Higgs
if $S_B$ mass is heavier than $h_{\rm SM}$. 

In SUSY case, there is a tree-level upper bound on the $S_B$ mass,
$m_{S_B} \leq M_{Z_B}$, similarly to the bound on the 
neutral Higgs mass $m_h \leq M_Z$ at tree level. 
This upper bound however can be raised somewhat by loop effects 
involving squarks 
(especially scalar mirror quarks). 
Therefore $S_B$ cannot be too heavy and $S_B \rightarrow Z_B Z_B$ is 
likely to be kinematically forbidden. Its main decay
will be into the SM fermions or weak gauge bosons through 
$S_B - h_{\rm SM}$ mixing induced by the one loop involving 
squarks.
Again the final states are 
$Z_B S_B \rightarrow (q\bar{q}) (b\bar{b}, t\bar{t}, WW, ZZ)$.
The strategy for searching $S_B$ would be similar to Higgs boson search, but 
is probably more difficult if $Z_B \rightarrow q\bar{q}$. On the other hand, 
our model has a light CDM, and $Z_B$ has a moderate invisible branching ratio 
$\sim 21 \%$.  Therefore this could be used to suppress the QCD background.

\section{Conclusions}

If we assume that the peak around $m_{jj} \sim 140$ GeV reported 
by CDF in the $l \nu jj$ channel is due to leptophobic $Z_B$, 
the reported cross section can be reproduced if $g_B \sim 0.8$.  
Within anomaly-free (non)supersymmetric  $U(1)_B \times U(1)_L$ model 
with baryonic and leptonic CDM candidates, we studied the implications 
on dark matter physics and other possible collider signatures.
In particular, SUSY $U(1)_B$ model predicts a light baryonic   
Dirac fermion CDM.  Its direct detection cross section is predicted 
to be in the range of $0.01 - 0.05$ pb, which is somewhat larger than the 
DAMA or CoGeNT region.
The CRESST experiment is subtle because it touches this region.
If the CDF dijet excess is confirmed with its present value,
it is inevitable to consider heavier CDM or sizable Yukawa contributions.
Then the relic density of our light CDM becomes subdominant and
the WMAP data could be explained by the other CDMs.
It would also be very important to study the collider 
signatures of our scenario.  In fact, the monojet signals at Tevatron 
and LHC will strongly constrain our scenario.

On the other hand, if this excess in dijet becomes less prominent 
in the future, 
$W Z_B$ or its relative modes will constrain the $U(1)_B$ sector in terms 
of $M_{Z_B}$ and $g_B$, and the implication for baryonic CDM will be modified. 
If the $g_B$ should be weaker than the value we adopted in this Letter, 
annihilation cross section of baryonic CDM studied in this Letter may not be 
large enough, and we may have to include other contributions such as 
mirror fermions or their superpartners, as well as scalar exchanges. 
In this case, we have a number of additional parameters in the Yukawa 
couplings involving mirror fermions, and mixings among scalar bosons,
and phenomenological analysis of the model would be very involved.   
It would be a subject in the future when the situation about
the CDF dijet excess is clarified by other experiments. And our discussions
are complete within our model when the CDF dijet excess survives.

\vspace{0.5cm}

{\it Note Added}

While we are finalizing this Letter, there appeared a couple of papers
\cite{Jung:2011ua,Buckley:2011vs}. 
%The scenario with smaller $g_B$ is described in the text. 
In Ref.~ \cite{Buckley:2011vs}, the authors consider a similar model to 
this work, concentrating on light baryonic scalar dark matter $X_B$. 
Their results on the direct detection of the scalar dark matter are similar 
to ours, and $\sigma_{SI}$ seems to be small enough to evade the bound 
from CRESST because of the small $g_B\sim 0.3$.
However in this case the contribution to the $Wjj$ production
is too small ($\sim 0.3$ pb). 
We emphasize that the coupling used in Ref.~\cite{Buckley:2011vs}
for explanation of CoGeNT/DAMA data cannot explain the CDF $Wjj$ excess.
In the present work we also discussed the model with
the fermionic dark matter and presented more careful discussion on
the collider signature including the monojet + missing energy signals
at the LHC.
The authors of Ref.~\cite{Liu:2011di} discussed the CDF dijet anomaly
within a $U(1)_X$ Stueckelberg extension of the SM. Effectively 
their model is almost same as our model except the coupling,
baryonic charge and dark matter contents.  

\section*{Acknowledgements}
\begin{acknowledgments}
We thank Suyong Choi, Manuel Drees, Dong Hee Kim, Soo Bong Kim,  
Sungwon Lee and Intae Yu for useful communications and suggestions.
This work is supported in part by National Research Foundation through 
Korea Neutrino Research Center (KNRC) at Seoul National University (PK).
The work of CY is supported by Basic Science Research Program
through NRF (2011-0022996).
\end{acknowledgments}


\end{document}